\documentclass[conference]{IEEEtran}

\usepackage{cite}

\ifCLASSINFOpdf
  \usepackage[pdftex]{graphicx}
\else
   \usepackage[dvips]{graphicx}
\fi

\usepackage[cmex10]{amsmath}
\usepackage{amssymb}

\usepackage{array}
\usepackage[ruled,boxed,linesnumbered]{algorithm2e}

\usepackage[caption=false, font=footnotesize]{subfig}

\usepackage{multirow}
\usepackage{threeparttable}
\usepackage{url}
\usepackage{stfloats} 
\newtheorem{theorem}{Theorem}

\begin{document}
%
\title{Reduced-Complexity Maximum-Likelihood Decoding for 3D MIMO Code}

\author{
\IEEEauthorblockN{Ming Liu, Jean-Fran\c{c}ois H\'elard, Matthieu Crussi\`ere, Maryline H\'elard}
\IEEEauthorblockA{Universit\'e Europ\'eenne de Bretagne (UEB)\\
INSA, IETR, UMR 6164, F-35708, Rennes, France \\
Email: \{ming.liu; jean-francois.helard; matthieu.crussiere;  maryline.helard\}@insa-rennes.fr}
}

\maketitle

\begin{abstract}
The 3D MIMO code is a robust and efficient space-time coding scheme for the distributed MIMO broadcasting. However, it suffers from the high computational complexity if the optimal maximum-likelihood (ML) decoding is used.
In this paper we first investigate the unique properties of the 3D MIMO code and consequently propose a simplified decoding algorithm without sacrificing the ML optimality.
Analysis shows that the decoding complexity is reduced from $O(M^8)$ to $O(M^{4.5})$ in quasi-static channels when $M$-ary square QAM constellation is used.
Moreover, we propose an efficient implementation of the simplified ML decoder which achieves a much lower decoding time delay compared to the classical sphere decoder with Schnorr-Euchner enumeration.
\end{abstract}
\IEEEpeerreviewmaketitle

%
\IEEEpeerreviewmaketitle

\section{Introduction}
Multiple-input multiple-output (MIMO) technology in combination with the space-time block code (STBC) offers increased spectral efficiency and improved reliability without requiring additional spectrum bandwidth~\cite{tarokh1998space}.
Hence, it has been widely adopted by many state-of-the-art communication systems such as IEEE 802.11n, 3GPP Long Term Evolution (LTE) and WiMAX etc.
It is also considered as a core technique for the future TV broadcasting system~\cite{DVB_NGH}.

A so-called Space-Time-Space (3D) MIMO code has been proposed for the $4\times 2$ distributed MIMO broadcasting in which MIMO modulated signal is sent from two cooperating transmission sites to the receivers in the coverage are. Each site has two transmit antennas and each receiver equips two receive antennas, as well~\cite{nasser20083d}.
The 3D MIMO code combines the robustness of Alamouti scheme~\cite{alamouti98simple} with the efficiency of the Golden code~\cite{belfiore2005golden}.
Hence it offers reliable performance even in presence of strong received signal power imbalances.
The latest study~\cite{ENGINES_D23} shows that the 3D MIMO code is the most efficient and robust in distributed MIMO broadcasting scenarios compared with other state-of-the-art STBCs such as DjABBA code~\cite{hotinen03multiantenna}, BHV code~\cite{biglieri09fast} and Srinath-Rajan code~\cite{srinath09low}, which suggests it is a promising candidate for the future distributed MIMO broadcasting systems.

However, as eight $M$-QAM modulated information symbols are stacked within one 3D MIMO codeword, the computational complexity is as high as $O(M^8)$ when maximum-likelihood (ML) decoding is adopted.
No study on reducing the ML decoding complexity of 3D MIMO code has been conducted in the literature.
\cite{polonen2011reduced} proposed another fast-decodable STBC for the distributed MIMO broadcasting. However, it does not achieve full-rate and its performance is not as good as 3D MIMO code.

In this work, we first reveal some unique properties of the 3D MIMO code which have not been presented in any previous work. Based on these properties, we propose a novel simplified ML detection method.
The complexity reduction is achieved by three means: 1) embedded orthogonality coming from Alamouti-like block structure in the codeword enables a group-wise detection; 2) embedded orthogonality between real and imaginary parts of the symbol group inheriting from the Golden code enables independent detection of real and imaginary parts in parallel; 3) adaptive searching radius avoids the cumbersome exhaustive search.

The reminder of the paper is organized as follows. The 3D MIMO code and the MIMO system model is introduced in Section~\ref{sec:model}.
The novel simplified ML decoding algorithm is proposed in Section~\ref{sec:algo}.
Complexity analysis and simulation results are presented in Section~\ref{sec:simu}.
Finally, conclusions are drawn in Section~\ref{sec:con}.

In this paper, $x^R$ and $x^I$ represent the real and imaginary parts of the complex number $x$, respectively.
The function $\check{x}$ performs the complex-real conversion as: $\check{x}\triangleq [[x^R\ x^I ]^T\ [-x^I\ x^R]^T]$.
For a complex vector $\mathbf{x}=[x_1,x_2,\ldots, x_n]^T$, the function $\widetilde{\mathbf{x}}$ separates the real and imaginary parts of the complex vector, i.e. $\widetilde{\mathbf{x}}\triangleq[x_1^R,x_1^I,\ldots, x_n^R,x_n^I]^T$.
For a matrix $\mathbf{X}=[\mathbf{x}_1,\mathbf{x}_2,\ldots, \mathbf{x}_n]$ where $\mathbf{x}_j$ is the $j$th column of $\mathbf{X}$, the function $vec(\mathbf{X})$ denotes stacking the columns of $\mathbf{X}$ one below another, i.e. $vec(\mathbf{X})\triangleq[\mathbf{x}_1^T,\mathbf{x}_2^T,\ldots, \mathbf{x}_n^T]^T$.
Consequently, $\widetilde{vec(\mathbf{X})}$ denotes vectorizing matrix $\mathbf{X}$ followed by the real/imaginary part separation.
The inner product of two vectors $\mathbf{x}$ and $\mathbf{y}$ is denoted by $\langle \mathbf{x},\mathbf{y}\rangle$.

\section{3D MIMO Code and System Model}
\label{sec:model}

The codeword matrix of the 3D MIMO code is given in (\ref{eq:3D}) which is shown on next page,
where $\theta=\frac{1+\sqrt{5}}{2}$, $\bar{\theta}=1-\theta$, $\alpha=1+i(1-\theta)$ and $\bar{\alpha}=1+i(1-\bar{\theta})$ with $i=\sqrt{-1}$.
The codeword is formed by arranging two Golden codewords $\mathbf{X}_1$ and $\mathbf{X}_2$ in an Alamouti manner.
It achieves full-diversity.
Since eight information symbols $\mathbf{s}=[s_1,\ldots, s_8]^T$ are transmitted over four ($T=4$) uses, it achieves a space-time (ST) coding rate of two which is full-rate for $4\times2$ MIMO transmission.
\newcounter{MYtempeqncnt1}
\begin{figure*}[!t]
\normalsize
\vspace*{4pt}
\setcounter{MYtempeqncnt1}{\value{equation}}
\setcounter{equation}{0}
\begin{equation}
\label{eq:3D}
\textbf{X}_{\mathrm{3D}}=\left [\begin{array}{*{20}c}
        \mathbf{X}_1 & -\mathbf{X}_2^{\ast}\\
        \mathbf{X}_2 & \mathbf{X}_1^{\ast} \\
        \end{array}\right]
=\!\frac{1}{\sqrt{5}}\!\left [\begin{smallmatrix}
        \alpha (s_1+\theta s_2) & \alpha (s_3+\theta s_4) & -\alpha^{\ast} (s_5^{\ast}+\theta s_6^{\ast}) & -\alpha^{\ast} (s_7^{\ast}+\theta s_8^{\ast})\\
        i\bar{\alpha} (s_3+\bar{\theta} s_4) & \bar{\alpha} (s_1+\bar{\theta} s_2)  & i\bar{\alpha}^{\ast} (s_7^{\ast}+\bar{\theta} s_8^{\ast}) & -\bar{\alpha}^{\ast} (s_5^{\ast}+\bar{\theta} s_6^{\ast}) \\
        \alpha (s_5+\theta s_6) & \alpha (s_7+\theta s_8) & \alpha^{\ast} (s_1^{\ast}+\theta s_2^{\ast}) & \alpha^{\ast} (s_3^{\ast}+\theta s_4^{\ast})\\
        i\bar{\alpha} (s_7+\bar{\theta} s_8) & \bar{\alpha} (s_5+\bar{\theta} s_6)  & -i\bar{\alpha}^{\ast} (s_3^{\ast}+\bar{\theta} s_4^{\ast}) & \bar{\alpha}^{\ast} (s_1^{\ast}+\bar{\theta} s_2^{\ast}) \\
\end{smallmatrix}
        \right]
\end{equation}
\setcounter{equation}{1}
\hrulefill
\end{figure*}

The 3D MIMO code is a linear STBC and its codeword matrix can be constructed as\cite{biglieri09fast}:
\begin{equation}\label{eq:ST_coding_by_generator_matrix}
    \widetilde{vec(\mathbf{X})} = \mathbf{G}\widetilde{\mathbf{s}},
\end{equation}
where $\mathbf{G}=[\widetilde{vec(\mathcal{A}_1)}, \widetilde{vec(\mathcal{B}_1)},\ldots,\widetilde{vec(\mathcal{B}_8)}]$ is the generator matrix with $\mathcal{A}_j$ and $\mathcal{B}_j$ being the the weight matrices representing the contribution of the real and imaginary parts of the $j$th information symbol $s_j$ in the final codeword matrix~\cite{srinath09low}.

For the MIMO system with $N_t$ transmit and $N_r$ receive antennas, the signal transmission over quasi-static flat-fading channel is expressed as:
\begin{equation}\label{eq:rec_sig}
    \mathbf{Y}=\mathbf{H}\mathbf{X}+\mathbf{N},
\end{equation}
where $\mathbf{Y}$ and $\mathbf{N}$ are $N_r\times T$ matrices representing received signal and complex-valued additive white Gaussian noise (AWGN) component, respectively;
$\mathbf{X}$ is an $N_t\times T$ matrix representing a codeword of the STBC;
$\mathbf{H}$ is an $N_r\times N_t$ matrix in which the $(j,k)$th element $h_{j,k}$ is the gain of the channel link between the $k$th transmit antenna and $j$th receive antenna.
Separating the real and imaginary parts and stacking the columns of the transmitted/received signal, it yields the signal expression in real-value form:
\begin{equation}\label{eq:rec_sig_real_eq}
    \widetilde{\mathbf{y}}=\mathbf{H}_{eq}\widetilde{\mathbf{s}}+\widetilde{\mathbf{n}},
\end{equation}
where $\widetilde{\mathbf{y}} = \widetilde{vec(\mathbf{Y})}$, $\widetilde{\mathbf{n}}=\widetilde{vec(\mathbf{N})}$ and $\mathbf{H}_{eq}$ is the equivalent channel matrix and is written as:
\begin{equation}\label{eq:Heq}
 \mathbf{H}_{eq}=(\mathbf{I}_{T}\otimes\check{\mathbf{H}})\mathbf{G},
\end{equation}
where $\mathbf{I}_{T}$ is the $T\times T$ identity matrix, $\otimes$ is the Kronecker product.

\section{Simplified ML Detection for 3D MIMO code}
\label{sec:algo}
\subsection{ML decoding of STBC}
Since the received signal $\widetilde{\mathbf{y}}$ in (\ref{eq:rec_sig_real_eq}) can be viewed as lattice points perturbed by the noise, the maximum-likelihood (ML) solution of the transmitted signal is the combination of the information symbol $\widetilde{\mathbf{s}}$ which has minimal Euclidean distance to the received signal $\widetilde{\mathbf{y}}$, namely:
\begin{equation}\label{eq:ML_detection}
    \hat{\mathbf{s}}^{\mathrm{ML}}=\arg\min_{\mathbf{s}\in\boldsymbol\Theta^8}\|\widetilde{\mathbf{y}}-\mathbf{H}_{eq}\widetilde{\mathbf{s}} \|^2,
\end{equation}
where $\boldsymbol\Theta$ is the set of the constellation of complex-valued information symbols, and $\boldsymbol\Theta^8$ indicates that the symbol vector $\mathbf{s}$ consists of eight independently selected constellation points.
It means that the optimal solution is found by jointly determining eight information symbols.
Specifically, when the $M$-QAM modulation is adopted by the information symbols, a brute-force searching of $\hat{\mathbf{s}}^{\mathrm{ML}}$ requires testing all $M^8$ possibilities of the signal vector, which is computationally intensive.

Fast decoding of the STBC based on orthogonal-triangular (QR) decomposition has been discussed in literatures~\cite{srinath09low,biglieri09fast,sinnokrot2010fast}.
More precisely, by performing Gram-Schmidt procedure to the columns of the equivalent channel matrix $\mathbf{H}_{eq}$, it yields an unitary matrix $\mathbf{Q}$ and an upper triangular matrix $\mathbf{R}$, i.e. $\mathbf{H}_{eq}=\mathbf{Q}\mathbf{R}$ where $\mathbf{Q}\triangleq[\mathbf{q}_1,\ldots,\mathbf{q}_{16}]$ and
\begin{equation}
   \mathbf{R}\triangleq\left [\begin{array}{*{20}c}
  \|\mathbf{r}_1\|^2 & \langle \mathbf{q}_1,\mathbf{h}_2\rangle & \cdots & \langle \mathbf{q}_1,\mathbf{h}_{16}\rangle \\
  0 & \|\mathbf{r}_2\|^2 & \cdots & \langle \mathbf{q}_2,\mathbf{h}_{16}\rangle \\
  \vdots & \vdots & \ddots & \vdots \\
  0&0 &\cdots & \|\mathbf{r}_{16}\|^2 \\
 \end{array}\right],
\end{equation}
where $\mathbf{r}_1=\mathbf{h}_1$, $\mathbf{r}_{j}=\mathbf{h}_j-\sum_{k=1}^{j-1}\langle \mathbf{q}_k,\mathbf{h}_j\rangle \mathbf{q}_k$,
$\mathbf{q}_j=\mathbf{r}_j/\|\mathbf{r}_j\|$, $j=1,\ldots,16$.

Instead of solving (\ref{eq:ML_detection}), the ML solution can be alternatively obtained by:
\begin{equation}\label{eq:SD_detection_metric}
    \hat{\mathbf{s}}^{\mathrm{ML}}=\arg\min_{\mathbf{s}\in\boldsymbol\Theta^8}\|\widetilde{\mathbf{z}}-\mathbf{R}\widetilde{\mathbf{s}}\|^2,
\end{equation}
where $\widetilde{\mathbf{z}}=\mathbf{Q}^{\mathcal{H}}\widetilde{\mathbf{y}}$ is the real-valued received signal after a linear operation $\mathbf{Q}^{\mathcal{H}}$.
For a well-designed STBC, some elements of $\mathbf{R}$ are equal to zero, which permits some information symbols to be determined independently from others. In other words, a joint detection in high dimension is turned to several independent detections in low dimension, leading to a significant reduction of decoding complexity~\cite{srinath09low,sinnokrot2010fast}.

\subsection{Important properties of 3D MIMO code}
With the definitions in (\ref{eq:3D}), (\ref{eq:ST_coding_by_generator_matrix}) and (\ref{eq:Heq}), we can derive the real-valued $16\times 16$ equivalent channel matrix $\mathbf{H}_{eq}$. 
Rewrite the upper triangular matrix $\mathbf{R}$:
\begin{equation}\label{eq:R_mat}
 \mathbf{R}=\left [\begin{array}{*{20}c}
  \mathbf{R}_{11} & \mathbf{R}_{12} & \mathbf{R}_{13}  & \mathbf{R}_{14} \\
  0& \mathbf{R}_{22} & \mathbf{R}_{23} & \mathbf{R}_{24} \\
  0 & 0& \mathbf{R}_{33} & \mathbf{R}_{34} \\
  0&0 &0 & \mathbf{R}_{44} \\
 \end{array}\right],
\end{equation}
where $\mathbf{R}_{jk}$'s are $4\times 4$ matrices containing $\langle \mathbf{q}_m,\mathbf{h}_n\rangle$'s with $m=4(j-1)+1,\ldots,4j$ and $n=4(k-1)+1,\ldots,4k$. More importantly, $\mathbf{R}$ matrix has the following interesting properties.

\vspace{.1cm}
\begin{theorem}
    \label{theorm:1}
  $\mathbf{R}_{11}$ is an upper triangular matrix with $\langle \mathbf{q}_1,\mathbf{h}_2\rangle=\langle \mathbf{q}_1,\mathbf{h}_4\rangle=\langle \mathbf{q}_2,\mathbf{h}_3\rangle=\langle \mathbf{q}_3,\mathbf{h}_4\rangle=0$.
\end{theorem}
\begin{IEEEproof}
According to the definition of QR decomposition, $\mathbf{R}_{11}$ is an upper triangular matrix.

    With some straightforward computations based on $\mathbf{H}_{eq}$, it yields $\langle \mathbf{h}_1,\mathbf{h}_2\rangle=\langle \mathbf{h}_1,\mathbf{h}_4\rangle=\langle \mathbf{h}_2,\mathbf{h}_3\rangle=\langle \mathbf{h}_3,\mathbf{h}_4\rangle=0$. From the definition of QR decomposition, $\mathbf{q}_1=\mathbf{h}_1/\|\mathbf{h}_1\|$. Hence, $\langle \mathbf{q}_1,\mathbf{h}_2\rangle=\langle \mathbf{q}_1,\mathbf{h}_4\rangle=0$.

    In addition, $\mathbf{r}_{2}=\mathbf{h}_2-\langle \mathbf{q}_1,\mathbf{h}_2\rangle \mathbf{q}_1=\mathbf{h}_2$, $\mathbf{q}_2=\mathbf{r}_2/\|\mathbf{r}_2\|= \mathbf{h}_2/\|\mathbf{h}_2\|$.
    Therefore, using $\langle \mathbf{h}_2,\mathbf{h}_3\rangle=0$, it yields $\langle \mathbf{q}_2,\mathbf{h}_3\rangle=0$.

    Finally, $\mathbf{r}_{3}=\mathbf{h}_3-\sum_{j=1}^{2}\langle \mathbf{q}_j,\mathbf{h}_3\rangle \mathbf{q}_j=\mathbf{h}_3-\langle \mathbf{q}_1,\mathbf{h}_3\rangle \mathbf{q}_1$, $\mathbf{q}_3=(\mathbf{h}_3-\langle \mathbf{q}_1,\mathbf{h}_3\rangle \mathbf{q}_1)/\|\mathbf{r}_3\|$. Therefore, $\langle \mathbf{q}_3,\mathbf{h}_4\rangle=(\langle \mathbf{h}_3,\mathbf{h}_4\rangle-\langle \mathbf{q}_1,\mathbf{h}_3\rangle\langle \mathbf{q}_1,\mathbf{h}_4\rangle)/\|\mathbf{r}_3\|=0$.
\end{IEEEproof}

\vspace{.1cm}
Remark: Theorem~\ref{theorm:1} suggests that the real and imaginary parts of information symbol $s_1$ and $s_2$ can be decoded independently.
Using similar idea and procedure, the same property can be derived for $\mathbf{R}_{jj}$, $j=2,3,4$. 

\vspace{.1cm}
\begin{theorem}
\label{theo:R13}
  $\mathbf{R}_{13}$ is a null matrix when the channel is quasi-static, i.e. $\langle \mathbf{q}_j,\mathbf{h}_k\rangle=0$ with $j=1,2,3,4$ and $k=9,10,11,12$.
\end{theorem}
\begin{IEEEproof}
Using the same method as in the proof of Theorem~\ref{theorm:1}, it yields $\langle \mathbf{q}_j,\mathbf{h}_k\rangle=0$, $\forall j=1,2,3$, $k=9,10,11,12$.

Taking into account that $\langle \mathbf{q}_1,\mathbf{h}_4\rangle=\langle \mathbf{q}_3,\mathbf{h}_4\rangle=0$, it yields $\mathbf{r}_{4}=\mathbf{h}_4-\sum_{j=1}^{3}\langle \mathbf{q}_j,\mathbf{h}_4\rangle \mathbf{q}_j= \mathbf{h}_4-\langle \mathbf{q}_2,\mathbf{h}_4\rangle \mathbf{q}_2$. Hence, $\langle \mathbf{q}_4,\mathbf{h}_k\rangle=(\langle\mathbf{h}_4,\mathbf{h}_k\rangle-\langle \mathbf{q}_2,\mathbf{h}_4\rangle \langle\mathbf{q}_2,\mathbf{h}_k\rangle)/\|\mathbf{r}_4\|=0$, $\forall k=9,10,11,12$.
\end{IEEEproof}
\vspace{.1cm}
Remark: Theorem~\ref{theo:R13} suggests that $z_1$ and $z_2$ do not contain contribution from $s_5$ and $s_6$.
It enables separating decoding into groups.
The orthogonalities between columns partially come from the Alamouti structure embedded in the codeword which requires the quasi-staticity of the channel.

\vspace{.1cm}
\begin{theorem}
\label{theo:R23_QR}
     Performing QR decomposition $\mathbf{R}_{23}= \mathbf{E}\mathbf{F}$, the yielding upper triangular matrix $\mathbf{F}$ has similar structure as $\mathbf{R}_{11}$, namely its $(1,2)$, $(1,4)$, $(2,3)$ and $(3,4)$ elements equal to zero.
\end{theorem}

\begin{IEEEproof}
Due to the length limitation, we omit the details of basic manipulations and only present some sketches of the proof.

Denote the $j$th column of $\mathbf{R}_{23}$ as $\mathbf{p}_j$, i.e. $\mathbf{R}_{23}=[\mathbf{p}_1,\mathbf{p}_2,\mathbf{p}_3,\mathbf{p}_4]$.
From the definition of $\mathbf{H}_{eq}$ and using previous two theorems, it is easy to prove that:
\begin{align}
 & \langle\mathbf{q}_{j},\mathbf{h}_{k}\rangle=\langle\mathbf{q}_{j+1},\mathbf{h}_{k+1}\rangle,\quad \forall j=5,7\ \mathrm{and}\ k=9,11, \label{eqn:qh1} \\
 & \langle\mathbf{q}_{j+1},\mathbf{h}_{k}\rangle=-\langle\mathbf{q}_{j},\mathbf{h}_{k+1}\rangle,\ \ \forall j=5,7\ \mathrm{and}\ k=9,11, \label{eqn:qh2} \\
 & \langle\mathbf{h}_{5},\mathbf{h}_{9}\rangle\langle\mathbf{h}_{6},\mathbf{h}_{11}\rangle
 -\langle\mathbf{h}_{5},\mathbf{h}_{11}\rangle\langle\mathbf{h}_{6},\mathbf{h}_{9}\rangle \nonumber\\
 &= \langle\mathbf{h}_{6},\mathbf{h}_{9}\rangle\langle\mathbf{h}_{7},\mathbf{h}_{9}\rangle -
 \langle\mathbf{h}_{5},\mathbf{h}_{9}\rangle\langle\mathbf{h}_{8},\mathbf{h}_{9}\rangle \nonumber\\
 &= \langle\mathbf{h}_{6},\mathbf{h}_{11}\rangle\langle\mathbf{h}_{7},\mathbf{h}_{9}\rangle
 -\langle\mathbf{h}_{5},\mathbf{h}_{11}\rangle\langle\mathbf{h}_{8},\mathbf{h}_{9}\rangle. \label{eqn:qh3}
\end{align}

(\ref{eqn:qh1}) and (\ref{eqn:qh2}) suggest that the first and the second columns of matrix $\mathbf{R}_{23}$ are orthogonal, i.e. $\langle \mathbf{p}_{1},\mathbf{p}_{2}\rangle=0$.
Hence, it is sufficient to assert that the $(1,2)$th element of matrix $\mathbf{F}$ is zero.

In addition, using the properties in (\ref{eqn:qh1}) and (\ref{eqn:qh2}), the inner product of matrix $\mathbf{R}_{23}$'s first and fourth columns writes:
\begin{align}\label{eq:p1p4}
    &\langle \mathbf{p}_1,\mathbf{p}_4\rangle = \langle \mathbf{q}_{6},\mathbf{h}_{9}\rangle\langle \mathbf{q}_{5},\mathbf{h}_{11}\rangle -\langle \mathbf{q}_{5},\mathbf{h}_{9}\rangle\langle \mathbf{q}_{6},\mathbf{h}_{11}\rangle  \nonumber \\
     &+ \langle \mathbf{q}_{8},\mathbf{h}_{9}\rangle\langle \mathbf{q}_{7},\mathbf{h}_{11}\rangle
     - \langle \mathbf{q}_{7},\mathbf{h}_{9}\rangle\langle \mathbf{q}_{8},\mathbf{h}_{11}\rangle \nonumber  \\
& =\frac{1}{\|\mathbf{r}_5\|^2}\Big(\langle\mathbf{h}_{5},\mathbf{h}_{11}\rangle\langle\mathbf{h}_{6},\mathbf{h}_{9}\rangle
 -\langle\mathbf{h}_{5},\mathbf{h}_{9}\rangle\langle\mathbf{h}_{6},\mathbf{h}_{11}\rangle\Big) \nonumber \\
 &+\frac{1}{\|\mathbf{r}_7\|^2}\Big[ \langle\mathbf{h}_{6},\mathbf{h}_{9}\rangle\langle\mathbf{h}_{7},\mathbf{h}_{9}\rangle
 -\langle\mathbf{h}_{5},\mathbf{h}_{9}\rangle\langle\mathbf{h}_{8},\mathbf{h}_{9}\rangle \nonumber\\
 &+\frac{1}{\|\mathbf{h}_5\|^2}\langle\mathbf{h}_{5},\mathbf{h}_{7}\rangle \big(\langle\mathbf{h}_{6},\mathbf{h}_{11}\rangle\langle\mathbf{h}_{7},\mathbf{h}_{9}\rangle
 -\langle\mathbf{h}_{5},\mathbf{h}_{11}\rangle\langle\mathbf{h}_{8},\mathbf{h}_{9}\rangle\big)\nonumber\\
 &-\frac{1}{\|\mathbf{h}_5\|^4}\langle\mathbf{h}_{5},\mathbf{h}_{7}\rangle^2 \big(\langle\mathbf{h}_{5},\mathbf{h}_{9}\rangle\langle\mathbf{h}_{6},\mathbf{h}_{11}\rangle
 -\langle\mathbf{h}_{5},\mathbf{h}_{11}\rangle\langle\mathbf{h}_{6},\mathbf{h}_{9}\rangle \big)\Big].
\end{align}
With the definition of $\mathbf{H}_{eq}$, it can be shown that:
\begin{equation}\label{eq:r7r5}
    \|\mathbf{r}_7\|^2=\|\mathbf{r}_5\|^2\Big(1+\frac{1}{\|\mathbf{h}_5\|^2}\langle\mathbf{h}_{5},\mathbf{h}_{7}\rangle-\frac{1}{\|\mathbf{h}_5\|^4}\langle\mathbf{h}_{5},\mathbf{h}_{7}\rangle^2\Big).
\end{equation}
Taking into account the equalities in (\ref{eqn:qh3}) and (\ref{eq:r7r5}), (\ref{eq:p1p4}) is turned to $\langle \mathbf{p}_1,\mathbf{p}_4\rangle = 0$, which is sufficient to assert that the $(1,4)$th element of matrix $\mathbf{F}$ is also zero.

Following similar procedure, we can prove that the $(2,3)$th and $(3,4)$th elements of $\mathbf{F}$ are equal to zero, as well.
\end{IEEEproof}
\vspace{.1cm}

\begin{figure}[!t]
\centering
\includegraphics[width=2.7in]{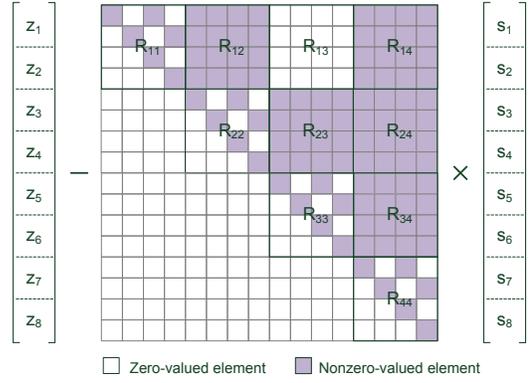}
\caption{Illustration of the ML decoding metric in quasi-static channel.}
\label{fig_R_mat}
\end{figure}

The aforementioned properties are illustrated in Fig.~\ref{fig_R_mat} where zero and nonzero entries of matrix $\mathbf{R}$ are easily seen.
These properties can be exploited to achieve low-complexity ML decoding, which will be demonstrated in the following parts.

\subsection{Simplified ML decoding}
Based on the theorems provided in the previous part, a simplified ML detection for the 3D MIMO code is proposed in this subsection.
The basic idea is that, using Theorems~\ref{theo:R13} and~\ref{theo:R23_QR}, the joint detection of eight information symbols is converted into several detections in lower searching dimension in parallel, which results in a lower global detection complexity. Moreover, using Theorem~\ref{theorm:1} and its inferences, the detection of complex information symbols is turned to independent detections of real and imaginary parts in parallel, which further reduces the complexity.

More precisely, with the knowledge of matrix $\mathbf{R}$ in (\ref{eq:R_mat}) and taking into account Theorem~\ref{theo:R13}, the ML detection metric in (\ref{eq:SD_detection_metric}) can be expressed as:
\begin{align}
\|\widetilde{\mathbf{z}}-\mathbf{R}\widetilde{\mathbf{s}} \|^2& = \|\mathbf{z}_{78}-\mathbf{R}_{44}\mathbf{d} \|^2 \label{eq:ML_metric_original1}\\
&+ \|\mathbf{z}_{56}-\mathbf{R}_{33}\mathbf{c}-\mathbf{R}_{34}\mathbf{d} \|^2 \label{eq:ML_metric_original2}  \\
&+ \|\mathbf{z}_{34}-\mathbf{R}_{22}\mathbf{b}-\mathbf{R}_{23}\mathbf{c}-\mathbf{R}_{24}\mathbf{d} \|^2 \label{eq:ML_metric_original3}  \\
&+ \|\mathbf{z}_{12}-\mathbf{R}_{11}\mathbf{a}-\mathbf{R}_{12}\mathbf{b}-\mathbf{R}_{14}\mathbf{d} \|^2, \label{eq:ML_metric_original4}
\end{align}
where $\mathbf{a}=\widetilde{[s_1, s_2]^T}$, $\mathbf{b}=\widetilde{[s_3, s_4]^T}$, $\mathbf{c}=\widetilde{[s_5, s_6]^T}$, $\mathbf{d}=\widetilde{[s_7, s_8]^T}$, $\mathbf{z}_{12}=\widetilde{[z_1, z_2]^T}$, $\mathbf{z}_{34}=\widetilde{[z_3, z_4]^T}$, $\mathbf{z}_{56}=\widetilde{[z_5, z_6]^T}$ and $\mathbf{z}_{78}=\widetilde{[z_7, z_8]^T}$.
From (\ref{eq:ML_metric_original1}) to (\ref{eq:ML_metric_original4}), it can be seen that symbol groups $\mathbf{a}$ and $\mathbf{c}$ can be determined independently from each other for given $\mathbf{b}$ and $\mathbf{d}$.
For instance, $\mathbf{a}$ is obtained by using only (\ref{eq:ML_metric_original4}), if $\mathbf{d}$ is already known. This motivates us to perform conditional detection~\cite{sirianunpiboon2010fast} to realize group-wise decoding.

Specifically, the ML detection (\ref{eq:SD_detection_metric}) can be rewritten in an equivalent form:
\begin{align}
\label{eq:min3}
\hat{\mathbf{s}}^{\mathrm{ML}}&=\arg\min_{[\mathbf{b},\mathbf{d}]\in\boldsymbol\Theta^4} \Big(\|\mathbf{z}_{78}-\mathbf{R}_{44}\mathbf{d} \|^2 \\
 &+ \arg\min_{\mathbf{a}\in\boldsymbol\Theta^2}\|\mathbf{v}_{12}-\mathbf{R}_{11}\mathbf{a} \|^2 \label{eq:min_a}\\
&+  \arg \min_{\mathbf{c}\in\boldsymbol\Theta^2} \big( \|\mathbf{v}_{56} - \mathbf{R}_{33}\mathbf{c}\|^2   +   \|\mathbf{v}_{34} - \mathbf{R}_{23}\mathbf{c} \|^2 \big)\!\Big)\label{eq:min_c},
\end{align}
where 
$\mathbf{v}_{56}=\mathbf{z}_{56}-\mathbf{R}_{34}\mathbf{d}$, $\mathbf{v}_{34} = \mathbf{z}_{34}-\mathbf{R}_{22}\mathbf{\mathbf{b}}-\mathbf{R}_{24}\mathbf{\mathbf{d}}$ and $\mathbf{v}_{12} = \mathbf{z}_{12}-\mathbf{R}_{12}\mathbf{\mathbf{b}}-\mathbf{R}_{14}\mathbf{\mathbf{d}}$.
It suggests that the joint searching of eight information symbols is turned into two independent searching of two information symbols (shown by the two minimum operations inside the parentheses) conditioned on the other four information symbols (the first minimum operation outside the parentheses). Therefore, the ML decoding complexity is reduced from $O(M^8)$ to $O(M^6)$. Note that this complexity reduction is achieved \textit{without} any constraint on the constellation of information symbols.

\begin{figure}[!t]
\centering
\includegraphics[width=2.7in]{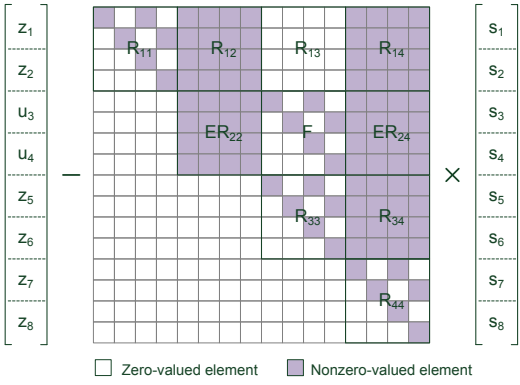}
\caption{Illustration of the modified ML decoding metric in quasi-static channel.}
\label{fig_R_mat_modified}
\end{figure}

In addition, Theorem~\ref{theorm:1} suggests that the real and imaginary parts of $\mathbf{a}$ can be determined independently, which results in a further reduction in complexity.
Interestingly, according to Theorem~\ref{theo:R23_QR}, the real and imaginary parts of $\mathbf{c}$ can be obtained independently, as well.
Specifically, the minimum operation for searching $\mathbf{c}$ in (\ref{eq:min_c}) is turned equivalently to:
\begin{equation}
\arg\min_{\mathbf{c}\in\boldsymbol\Theta^2}\Big(\|\mathbf{v}_{56}-\mathbf{R}_{33}\mathbf{c}\|^2 +  \|\mathbf{u}_{34}-\mathbf{F}\mathbf{c} \|^2\Big),
\end{equation}
where $\mathbf{u}_{34}=\mathbf{E}^T\mathbf{v}_{34}$.
The resulting ML detection metric is illustrated in Fig.~\ref{fig_R_mat_modified}, where the independency of the real and imaginary parts of $\mathbf{a}$ and $\mathbf{c}$ is clearly shown.
Provided that the real and imaginary parts of the information symbols are independently modulated (such as in the square QAM case), the ML decoding complexity is then reduced to $O(M^5)$.

Moreover, as far as the square QAM constellation is concerned, the decoding of real (imaginary) parts of information symbols can be further simplified. Take the detection of $\mathbf{a}$ as an example.
The square $M$-QAM complex symbols $\mathbf{a}$ are separated into $\sqrt{M}$-PAM real symbols on both real and imaginary axes, denoted as $\mathbf{a}^R=[s_1^R,s_2^R]^T$ and $\mathbf{a}^I=[s_1^I,s_2^I]^T$, respectively. The searching for $\mathbf{a}$ in (\ref{eq:min_a}) is converted into~\cite{sinnokrot2010fast}:
\begin{align}\label{eq:}
&\arg\min_{\mathbf{a}\in\boldsymbol\Theta^2}\|\mathbf{v}_{12}-\mathbf{R}_{11}\mathbf{a} \|^2= \nonumber \\
    & \arg\!\!\min_{\mathbf{a}^R\in\boldsymbol\Psi^2}\!\!\|\mathbf{v}_{12}^R-\mathbf{R}_{11}^R\mathbf{a}^R \|^2 \! + \!    \arg\!\!\min_{\mathbf{a}^I\in\boldsymbol\Psi^2}\!\!\|\mathbf{v}_{12}^I-\mathbf{R}_{11}^I\mathbf{a}^I \|^2,
\end{align}
where we slightly abuse the notation by denoting $\mathbf{v}_{12}^R$ ($\mathbf{v}_{12}^I$) as the first and third (second and fourth) elements of $\mathbf{v}_{12}$, $\mathbf{R}_{11}^R$ ($\mathbf{R}_{11}^I$) is tailored accordingly, $\boldsymbol\Psi$ is the set of $\sqrt{M}$-PAM constellation points.
Furthermore, the conditional detection is applied again here.
For a given $s_2^R$, the metric of the real part writes:
\begin{equation}\label{eq:}
    \|\mathbf{v}_{12}^R-\mathbf{R}_{11}^R\mathbf{a}^R \|^2\! = \! \big(\mathbf{w}_{12}(1)-\mathbf{R}_{11}(1,1)s_1^R\big)^2 + \mathbf{w}_{12}(3)^2,
\end{equation}
where $\mathbf{w}_{12}(1)=\mathbf{v}_{12}(1)-\mathbf{R}_{11}(1,3)s_2 ^R$ and $\mathbf{w}_{12}(3)=\mathbf{v}_{12}(3)-\mathbf{R}_{11}(3,3)s_2 ^R$. The metric is a quadratic function of $s_1^R$. Therefore, the best PAM symbol that minimizes the metric is easily found by:
\begin{equation}\label{eq:s_1R}
    \hat{s}_1^R=\texttt{Q}\Big(\frac{\mathbf{v}_{12}(1)-\mathbf{R}_{11}(1,3)s_2^R}{\mathbf{R}_{11}(1,1)}\Big),
\end{equation}
where $\texttt{Q}(\cdot)$ is the slicing operation providing the PAM symbol that is closest to the given value.
Following the same procedure, for a given $s_6^R$ we have:
\begin{equation}\label{eq:s_5R}
    \hat{s}_5^R=\texttt{Q}\Big(\frac{\mathbf{w}_{56}(1)\mathbf{R}_{33}(1,1)+\mathbf{w}_{34}(1)\mathbf{F}(1,1)}{\mathbf{R}_{33}(1,1)^2 + \mathbf{F}(1,1)^2}\Big),
\end{equation}
where $\mathbf{w}_{56}(1)=\mathbf{v}_{56}(1)-\mathbf{R}_{33}(1,3)s_6^R$ and $\mathbf{w}_{34}(1)=\mathbf{u}_{34}(1)-\mathbf{F}(1,3)s_6^R$.
Similar expressions can be derived for $\hat{s}_1^I$ and $\hat{s}_5^I$, as well.
Obviously, decoding of real symbol groups such as $\mathbf{a}^R$ is turned into two searchings over $\sqrt{M}$-PAM constellation points requiring a complexity of $O(\sqrt{M})$. Therefore, the decoding complexity is reduced to $O(M^{4.5})$.
\textit{Eventually, fully exploiting all the aforementioned properties, the 3D MIMO code turns out to be a fast decodable STBC.}

The ML decoding complexities of state-of-the-art $4\times 2$ rate-2 STBCs are compared in Table~\ref{tbl:complexity}. Note that all these STBCs require the quasi-staticity of the channel to achieve the claimed low complexities.
It can be seen that the 3D MIMO code requires equivalent complexity as other fast-decodable STBCs.

\begin{table}[!t]
\renewcommand{\arraystretch}{1.1}
\caption{Comparison of ML decoding complexities of STBCs}
\label{tbl:complexity}
\centering
\begin{tabular}{|c|c|c|}
\hline
\multirow{2}{*}{\textbf{STBC}} & \multicolumn{2}{c|}{\textbf{ML decoding complexity}}\\ \cline{2-3}
& any QAM & square QAM \\ \hline
3D MIMO~\cite{nasser20083d} & $O(M^6)$ & $O(M^{4.5})$\\ \hline
DjABBA~\cite{hotinen03multiantenna} & $O(M^7)$ & $O(M^{6})$\\ \hline
BHV~\cite{biglieri09fast} & $O(M^6)$ & $O(M^{4.5})$\\ \hline
Srinath-Rajan~\cite{srinath09low} & $O(M^5)$ & $O(M^{4.5})$\\ \hline
\end{tabular}
\end{table}

\subsection{Efficient implementation of the simplified ML decoder}

The pseudocode of an implementation of the simplified ML decoder is illustrated in Algorithm~\ref{algo_simplified} which is presented at the end of this paper.
Its major part follows the derivation in the previous subsection.
The two outermost `for' loops performs the traversal over combinations of $\mathbf{b}$ and $\mathbf{d}$.
A sorting function (denoted as $\texttt{sort}(\cdot)$ in line~\ref{algo_sort}) is used to arrange the possible combinations of $\mathbf{d}$ in ascending order with respect to
its distance from received signal.
It enables the early termination of the searching (in line~\ref{algo_78break}) once the distance resulted from the current $\mathbf{d}$ is greater than the minimum distance found in the previous searching.
The sorted set of $\mathbf{d}$ is denoted as $\bar{\boldsymbol\Theta}^2$.

The detection of $\mathbf{a}^R$, $\mathbf{a}^I$, $\mathbf{c}^R$ and $\mathbf{c}^I$ is implemented by real-valued sphere decoder~\cite{agrell2002closest}, as shown e.g. from
line \ref{algo_s2I_start} to \ref{algo_s2I_stop}  (or from
line \ref{algo_s6I_start} to \ref{algo_s6I_stop}).
Moreover, the Schnorr-Euchner (S-E) enumeration arranges the searching sequence according to the distances between the constellation points and the received signal to speed up the searching convergence~\cite{agrell2002closest}.
It can be simply implemented by look-up table $\texttt{S-E}(x)$ where $x$ is the  zero-forcing (ZF) result of the received signal (lines~\ref{algo_s2I_start} and \ref{algo_s6I_start}~)~\cite{wiesel2003efficient}. Note that the $\texttt{sort}$ function is actually implemented using the same technique to reduce the complexity.

Once a combination of information symbols having a smaller distance than the minimum distance in the previous search is found, the current solution $\mathbf{x}$ and the minimum distance $d_{min}$ are updated (line \ref{algo_updated}).
In other words, the searching radius is adaptively adjusted in the decoding progress, helping the fast convergence of the searching (lines~\ref{algo_termi_1}, \ref{algo_termi_2} and \ref{algo_termi_3}).

Note that Algorithm~\ref{algo_simplified} is a straightforward implementation of the proposed simplified decoder without sacrificing ML optimality.
Other techniques such as statistical tree pruning~\cite{ghaderipoor2008statistical} and sorted QR decomposition~\cite{wubben2001efficient} can also be incorporated in the implementation providing various performance-complexity trade-offs.

\section{Simulation}
\label{sec:simu}

\begin{figure}[!t]
\centering
\includegraphics[width=3in]{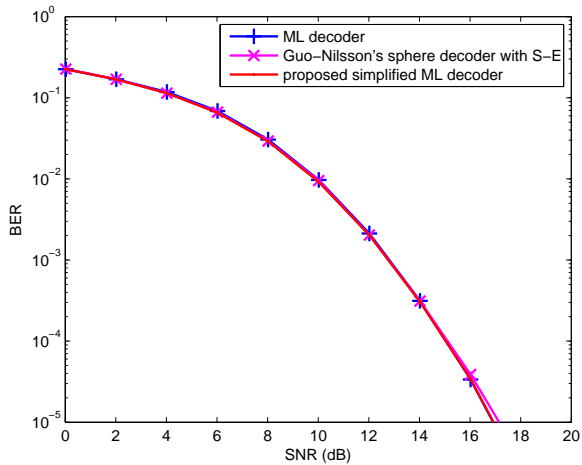}
\caption{BER comparison of sphere decoder with S-E, proposed simplified ML decoder and the ML decoder in quasi-static Rayleigh channel with 4-QAM.}
\label{fig_BER_comparison}
\end{figure}

\begin{figure}[!t]
\centering
\includegraphics[width=3in]{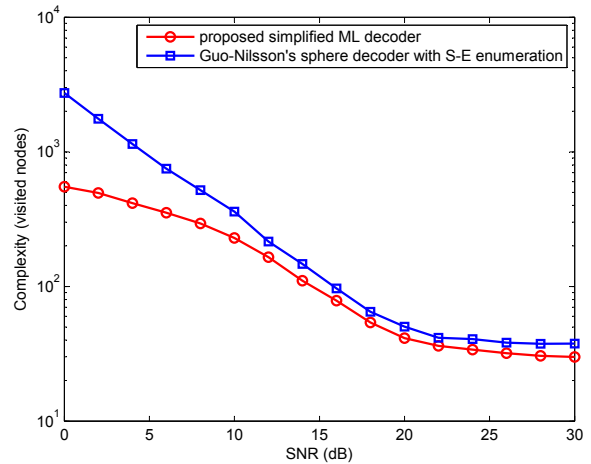}
\caption{Computational complexity required by sphere decoder with S-E and proposed simplified ML decoder, in quasi-static Rayleigh channel with 4-QAM constellation.}
\label{fig_complexity_comparison}
\end{figure}

We evaluate the proposed low-complexity ML decoder by simulation in this section.
The proposed decoder is implemented according to the pseudocode in Algorithm~\ref{algo_simplified}.
The sphere decoder with S-E enumeration is realized based on Guo-Nilsson's algorithm~\cite{guo2004reduced} which is an improved algorithm of the classical implementation~\cite{agrell2002closest} achieving much lower complexity than the original version.
As the proposed decoder contains four parallel searching branches, the processing time delay is determined by the \emph{maximum} visited nodes among all searching branches. In contrast, since the classical sphere decoder does not exploit the embedded properties of the code and follows a serial implementation, the delay is the time spent by the whole decoding process.
The comparison takes the common assumption that the processing time for checking each possible solution (referred to as `node') is approximately the same for both methods.
The channel is modeled as quasi-static i.i.d. Rayleigh fading channel. Symbol constellation is 4-QAM.

Fig.~\ref{fig_BER_comparison} presents the bit error rate (BER) performance of the proposed simplified decoder, Guo-Nilsson's sphere decoder and the optimal ML decoder without taking into account the channel coding.
This is to show the `pure' decoding performance of the STBC decoders.
The three decoders achieve almost the same performance.
Especially, the curve of proposed decoder overlaps with that of ML decoder, which suggests that the proposed simplified decoder provides the optimal decoding performance.

Fig.~\ref{fig_complexity_comparison} presents the decoding complexity in terms of processing time delay.
As can be seen from the figure, both decoders spend much less complexity than the ML decoder which needs to check $4^8=65536$ times.
Moreover, the proposed  decoder achieves a lower complexity than the classical sphere decoder within the whole signal-to-noise ratio (SNR) range.
Especially, the improvement is more significant in low SNR region.
For instance, the average visited nodes is reduced from $2738.9$ to $550.7$ at SNR of $0$ dB, namely about $80\%$ reduction in processing time.
The time reduction is over $53\%$ at SNR of $10$ dB.
The improvement decreases in higher SNR region i.e. $15$$\sim$$30$ dB. It is due to the fact that the ZF solution is more accurate at higher SNR and hence the S-E enumeration helps greatly improving the sphere decoding speed.
Nevertheless, in high SNR region, the nodes visited by the proposed method approaches to $29$ which is less than $37.6$, the amount required by the classical sphere decoder.
This still leads to $23\%$ reduction of processing time.

\section{Conclusion}
\label{sec:con}
In this work, we first explore some interesting properties of the 3D MIMO code. With this knowledge, we propose a simplified ML decoder which reduces the decoding complexity from $O(M^8)$ to $O(M^{4.5})$ in quasi-static channel.
Consequently we propose an implementation of the simplified ML decoder.
Simulation results show that the proposed simplified decoder needs less processing time, especially in the low SNR region, than the classical sphere decoder with S-E enumeration without sacrificing the ML decoding optimality.

\section*{Acknowledgment}

The authors would like to thank the support of French ANR project ``Mobile Multi-Media (M3)'' and ``P\^ole Images \& R\'eseaux''.

\IncMargin{.5em}
\begin{algorithm}[h]
 \SetAlgoLined
 \LinesNumbered
 \SetKwFunction{QR}{QR}
 \SetKwFunction{sort}{sort}
 \SetKwFunction{enum}{S-E}
 \SetKwFunction{slicing}{Q}
 \SetKwInOut{Input}{$\mathbf{y}$, $\mathbf{H}$}
 \SetKwInOut{Output}{output}
 $[\mathbf{Q},\mathbf{R}]$ = \QR{$\mathbf{H}_{eq}$}, $[\mathbf{E},\mathbf{F}]$ = \QR{$\mathbf{R}_{23}$}\;
 $\widetilde{\mathbf{z}}=\mathbf{Q}^T\widetilde{\mathbf{y}}$\;
 $d_{min}=\infty$\;
 $[\bar{\boldsymbol\Theta}^2,\ \bar{\boldsymbol\varepsilon}_{78}]$=\sort{$\boldsymbol\varepsilon_{78}=\| \mathbf{z}_{78} -\mathbf{R}_{44}\mathbf{d}\|^2,\ \forall \mathbf{d}$}\label{algo_sort}\;
 \For{$i= 1$ \KwTo $M^2$}{
    $\mathbf{d}=\bar{\boldsymbol\Theta}^2(i)$, compute $\mathbf{v}_{56}$\;
    \If{$\bar{\boldsymbol\varepsilon}_{78}(i)>d_{min}$\label{algo_termi_1}}{
        \textbf{break} \label{algo_78break}
    }
    \For{$l= 1$ \KwTo $M^2$}{
        $\mathbf{b}=\bar{\boldsymbol\Theta}^2(l)$,
        compute $\mathbf{v}_{12}$, $\mathbf{u}_{34}$ \;
        $\tau_{12}^R=\tau_{12}^I=\tau_{56}^R=\tau_{56}^I=\infty$\;
        $\overline{\boldsymbol\Psi}^2_2$ = \enum{$v_{2}^{R}/\mathbf{R}_{11}(3,3)$}\label{algo_s2I_start} \;
        \For{$k=1$ \KwTo $\sqrt{M}$}{
            $\hat{s}_2^R=\overline{\boldsymbol\Psi}^2_2(k)$\;
            $\varepsilon_2^R=|v_2^R-\mathbf{R}_{11}(3,3)\hat{s}_2^R|^2$\;
            \If{($\bar{\boldsymbol\varepsilon}_{78}(i)+\varepsilon_2^R)>d_{min}$\label{algo_termi_2}}
                {\textbf{break}}
            $\hat{s}_1^R$ = \slicing{$(v_1^{R}-\mathbf{R}_{11}(1,3)\hat{s}_2^R)/(\mathbf{R}_{11}(1,1))$}\;
            $\varepsilon_{12}^R=|v_1^R-\mathbf{R}_{11}(1,1)\hat{s}_1^R-\mathbf{R}_{11}(1,3)\hat{s}_2^R|^2+\varepsilon_2^R$\;
            \If{$\varepsilon_{12}^R<\tau_{12}^R$}{
                $x_1^R = \hat{s}_1^R$, $x_2^R = \hat{s}_2^R$, $\tau_{12}^R=\varepsilon_{12}^R$
            }
        }\label{algo_s2I_stop}
        run similar process as line \ref{algo_s2I_start} to \ref{algo_s2I_stop} for $s_1^I,\ s_2^I$

        $\overline{\boldsymbol\Psi}^2_6$ = \enum{$(\mathbf{R}_{33}(3,3)v_{6}^{R}+\mathbf{F}(3,3)u_{4}^{R})/(\mathbf{R}_{33}(3,3)^2+\mathbf{F}(3,3)^2)$} \label{algo_s6I_start} \;
        \For{$k=1$ \KwTo $\sqrt{M}$}{
            $\hat{s}_6^R=\overline{\boldsymbol\Psi}_6^2(k)$\;
            $\varepsilon_6^R=|v_6^R-\mathbf{R}_{33}(3,3)\hat{s}_6^R|^2+|u_4^R-\mathbf{F}(3,3)\hat{s}_6^R|^2$\;
            \If{($\bar{\boldsymbol\varepsilon}_{78}(i)+\varepsilon_6^R)>d_{min}$\label{algo_termi_3}}{
                \textbf{break}
                }
            $\hat{s}_5^R$ = \slicing{$((v_5^{R}-\mathbf{R}_{33}(1,3)\hat{s}_6^R)\mathbf{R}_{33}(1,1)+(u_3^{R}-\mathbf{F}(1,3)\hat{s}_6^R)\mathbf{F}(1,1))/(\mathbf{R}_{33}(1,1)^2+\mathbf{F}(1,1)^2)$}\;
            $\varepsilon_{56}^R=|v_5^R-\mathbf{R}_{33}(1,1)\hat{s}_5^R-\mathbf{R}_{33}(1,3)\hat{s}_6^R|^2+|u_3^R-\mathbf{F}(1,1)\hat{s}_5^R-\mathbf{F}(1,3)\hat{s}_6^R|^2+\varepsilon_6^R$\;
            \If{$\varepsilon_{56}^R<\tau_{56}^R$}{
                $x_5^R = \hat{s}_5^R$, $x_6^R = \hat{s}_6^R$, $\tau_{56}^R=\varepsilon_{56}^R$
            }
        }\label{algo_s6I_stop}

        run similar process as line \ref{algo_s6I_start} to \ref{algo_s6I_stop} for $s_5^I,\ s_6^I$
        $\tau = \tau_{12}^R+\tau_{12}^I+\tau_{56}^R+\tau_{56}^I+\bar{\boldsymbol\varepsilon}_{78}(i)$\;
        \If{$\tau<d_{min}$}{
            $\mathbf{x}=[x_1,x_2,\mathbf{b}^T,x_5,x_6,\mathbf{d}^T]^T$;  $d_{min}=\tau$ \label{algo_updated}\;
        }
    }

 }
 \caption{Implementation of the proposed simplified ML decoder for 3D MIMO code.}\label{algo_simplified}
\end{algorithm}
\DecMargin{.5em}

\bibliographystyle{IEEEtran}
\bibliography{WCNC2013}

\end{document}